\documentclass[10pt]{article}
\usepackage[final]{graphics}
\usepackage{amsfonts,amsbsy}

\def\empile#1\over#2{\mathrel{\mathop{\kern 0pt#1}\limits_{#2}}}

\newcommand{\sll}{\raise.15ex\hbox{$/$}\kern-.43em\hbox{$l$}}
\newcommand{\slepsilon}{\raise.15ex\hbox{$/$}\kern-.53em\hbox{$\epsilon$}}
\newcommand{\slvarepsilon}{\raise.15ex\hbox{$/$}\kern-.53em\hbox{$\varepsilon$}}
\newcommand{\slL}{\raise.15ex\hbox{$/$}\kern-.53em\hbox{$L$}}
\newcommand{\slP}{\raise.15ex\hbox{$/$}\kern-.53em\hbox{$P$}}
\newcommand{\slp}{\raise.1ex\hbox{$/$}\kern-.63em\hbox{$p$}}
\newcommand{\slq}{\raise.1ex\hbox{$/$}\kern-.53em\hbox{$q$}}
\newcommand{\slv}{\raise.1ex\hbox{$/$}\kern-.63em\hbox{$v$}}
\newcommand{\slR}{\raise.15ex\hbox{$/$}\kern-.53em\hbox{$R$}}
\newcommand{\slQ}{\raise.15ex\hbox{$/$}\kern-.53em\hbox{$Q$}}
\newcommand{\slK}{\raise.15ex\hbox{$/$}\kern-.53em\hbox{$K$}}
\newcommand{\slk}{\raise.15ex\hbox{$/$}\kern-.53em\hbox{$k$}}
\newcommand{\slSigma}{\raise.15ex\hbox{$/$}\kern-.53em\hbox{$\Sigma$}}
\newcommand{\slcalP}{\raise.15ex\hbox{$/$}\kern-.63em\hbox{$\cal P$}}
\newcommand{\slA}{\raise.15ex\hbox{$/$}\kern-.73em\hbox{$A$}}
\newcommand{\slbfA}{\raise.15ex\hbox{$/$}\kern-.73em\hbox{${\imb A}$}}
\newcommand{\slpartial}{\raise.15ex\hbox{$/$}\kern-.53em\hbox{$\partial$}}
\newcommand{\sla}{\raise.15ex\hbox{$/$}\kern-.53em\hbox{$a$}}
\newcommand{\slb}{\raise.15ex\hbox{$/$}\kern-.53em\hbox{$b$}}
\newcommand{\slc}{\raise.15ex\hbox{$/$}\kern-.53em\hbox{$c$}}

\font\tenimbf=cmmib10 at 10pt
\font\sevenimbf=cmmib10 at 7pt
\font\fiveimbf=cmmib10 at 5pt
\newfam\imbf
\textfont\imbf=\tenimbf
\scriptfont\imbf=\sevenimbf
\scriptscriptfont\imbf=\fiveimbf
\def\imb{\fam\imbf\tenimbf}

\def\p{{\boldsymbol p}}
\def\q{{\boldsymbol q}}

\def\k{{\boldsymbol k}}

\def\x{{\boldsymbol x}}
\def\y{{\boldsymbol y}}

\def\r{{\boldsymbol r}}

\def\b{{\boldsymbol b}}
\def\a{{\boldsymbol a}}
\def\cc{{\boldsymbol c}}

\def\be{\begin{eqnarray}}
\def\ee{\end{eqnarray}}

\begin{document}

\thispagestyle{empty}
\title {\bf Large mass $q\bar{q}$  production\\
 from the Color Glass Condensate}

\author{Fran\c cois Gelis$^{(1)}$ and Raju Venugopalan$^{(2)}$}
\maketitle
\begin{center}
\begin{enumerate}
\item Service de Physique Th\'eorique\\
  B\^at. 774, CEA/DSM/Saclay\\
  91191, Gif-sur-Yvette Cedex, France
\item Physics Department\\
  Brookhaven National Laboratory\\
  Upton, NY 11973, USA
\end{enumerate}
\end{center}

\begin{abstract}

\noindent We compute quark-antiquark pair production in the context of
the Color Glass Condensate model for central heavy-ion collisions.
The calculation is performed analytically to leading order in the
density of hard sources present in the projectiles, and is applicable
to quarks with a mass large compared to the saturation momentum. The
formulas derived in this paper are compared to expressions derived in
the framework of collinearly factorized perturbative QCD and in
$k_\perp$ factorization models. We comment on the breaking of
$k_\perp$ factorization which occurs beyond leading order in our
approach.

\end{abstract}

\section{Introduction}

Heavy Quark pair production in the collinear factorization formalism
of perturbative QCD is a well developed
subject~\cite{NasonDE1,NasonDE2,FrixiMNR1}. It is however not evident
that this formalism is applicable in the kinematic regime where the
heavy quark mass $m$ is much smaller than the center of mass energy
($m\ll\sqrt{s}$). In this kinematic regime, small $x$ effects may be
important.  An alternative formalism, the $k_\perp$ factorization
formalism, was developed to describe the physics in this
region~\cite{ColliE1,CatanCH1}. A key feature of this formalism is
that, unlike the collinear factorization formalism, the small $x$
gluons that produce the heavy quark pair have intrinsic transverse
momenta on the order of the hard scale of the process. At high $k_\perp$,
the collinear factorization formalism is recovered. A comparison of
the $k_\perp$ factorized formalism to recent hadroproduction data has been
performed in Ref.~\cite{HagleKSST1}.

In Ref.~\cite{LevinRSS1}, it was suggested that at small $x$, the
intrinsic transverse momenta in the $k_\perp$ factorized formalism are of
the order of the saturation scale $Q_s (x)$~\cite{GriboLR1,MuellQ1}.
Since the saturation scale grows with energy, so does the intrinsic
transverse momentum of gluons, thereby significantly increasing the
hadroproduction cross-section. The saturation inspired $k_\perp$
factorized formalism and phenomenological applications including
comparisons with the NLO perturbative QCD formalism are reviewed in
Ref.~\cite{RyskiSS1}.  A related approach is the color dipole approach
which has also been applied to heavy quark
hadroproduction~\cite{RaufeP1,KopelT1}.

A systematic way to study high parton density effects in QCD is the
Color Glass Condensate
(CGC)~\cite{McLerV1,McLerV2,McLerV3,JalilKLW1,JalilKLW2,JalilKLW3,JalilKLW4,KovneM1,KovneMW3,Balit1,Kovch1,Kovch3,JalilKMW1,IancuLM1,IancuLM2,IancuV1,IancuLM3,Muell4}.
Heavy quark production from the CGC was first discussed in the context
of deep inelastic scattering~\cite{McLerV4} and subsequently in the
context of inclusive and diffractive
photoproduction~\cite{GelisP1,GelisP2}. We will discuss here the
problem of hadro-production of heavy quark pairs at very high
energies. We will discuss specifically heavy quark production in the
scattering of two large nuclei which has some simplifying features.
Inclusive gluon production in the scattering of two nuclei has been
studied previously, first to lowest order in $\alpha_s$ and lowest
order in the parton density \cite{KovneMW1,KovneMW2,KovchR1,GyulaM1}
and later to all orders in the parton density (and lowest order in
$\alpha_s$)~\cite{KrasnV1,KrasnV2,KrasnNV2,Lappi1}.

The results for inclusive gluon production in the CGC approach, at
lowest order in the parton density, can be expressed in a $k_\perp$
factorized form~\cite{KovchR1,GyulaM1}. This $k_\perp$ factorization
is however broken when inclusive gluon production is computed to all
orders in the parton
density~\cite{KrasnV1,KrasnV2,KrasnNV2,Lappi1}. We expect a similar
breakdown of $k_\perp$ factorization in heavy quark production at the
small transverse momenta $\Lambda_{QCD} < k_\perp < Q_s$ when parton
density effects are computed to all orders~\cite{GelisKL1}. For a
recent discussion of $k_\perp$ factorization and the CGC, see
Ref.~\cite{GoncaM1}.

We do not explicitly include the effects of quantum evolution in our
approach -- the energy dependence of our results is determined entirely
by the $x$ dependence of the saturation scale $Q_s$. An interesting
consequence of quantum evolution in the CGC is the geometrical
scaling~\cite{StastGK1} of distributions with $Q_s$. This persists in
the kinematic window $Q_s < k_\perp < Q_s^2/\Lambda_{QCD}$ outside the
saturation region $k_\perp <
Q_s$~\cite{IancuIM1,IancuIM2,MuellT1,Trian1}. A number of
phenomenological models have argued that data at HERA for structure
functions, vector mesons and even
charm~\cite{StastGK1,KowalT1,MunieSM1,GoncaM2} (as well as data from
RHIC~\cite{SchafKMV1} and NMC~\cite{FreunRWS1}) exhibit geometrical
scaling. The competition between $k_\perp$ factorization preserving
quantum evolution effects and multiple-scattering $k_\perp$
factorization breaking effects immediately outside the saturation
regime is a very interesting
problem~\cite{KharzKT1,KharzLM1,JalilNV1,BaierKW1,AlbacAKSW1} but we will not
address it further here.

This paper is organized as follows. We begin in section 2.1 by
discussing the general formalism for pair production in a
time-dependent external field. The pair production probability for a
single pair is expressed in terms of the classical background field
produced by the color charge sources in the two nuclei. The
cross-section and the average number of pairs are obtained by
averaging the pair production probability with the appropriate weight
functional which governs the likelihood of different orientations of
the color charge density. In section 2.2, we argue that the leading
order contribution to pair production is of order O($\rho_1^2
\rho_2^2$), where $\rho_1$ and $\rho_2$ are the color charge densities
of the two sources. We write down explicit expressions, in covariant
gauge, for the classical field that contributes to the probability at
this order.  These covariant gauge expressions were first derived by
Kovchegov and Rischke~\cite{KovchR1}. For completeness, we reproduce
their derivation in appendix A. With these expressions for the
classical field, we write down an expression for the pair production
amplitude in section 2.3. In the following sub-section, we show that
this amplitude satisfies a Ward identity. The pair production
probability is computed in section 3. Some details of the computation
are given in appendix B. In section 4, we relate the results of our
computation to the $k_\perp$ factorization approach.  The breakdown of
$k_\perp$ factorization is discussed in the following section. We
conclude with a summary of our results.

\section{Pair production amplitude}
\subsection{Basics}
In order to compute the pair production amplitude in the Color Glass
Condensate model, we first recall some results~\cite{BaltzGMP1} about
particle production in an external time-dependent\footnote{More
exactly, one needs that the external field be time dependent {\sl in
any Lorentz frame}. For instance, a single moving nucleus, which
generates a time dependent classical color field in the laboratory
frame, cannot produce pairs.}  classical field. We denote
$G_{_{F}}(x,y)$ the Feynman (aka time-ordered) propagator of a quark
from the point $y$ to the point\footnote{The order of the points may
seem unnatural, but it is chosen so that the Dirac matrices (which
should be read from the endpoint to the starting point of the
propagator) appear with the correct ordering.} $x$ in the presence of
the external color field, and $G_{_{R}}(x,y)$ the retarded propagator
of the quark between the same points. Both are Green's functions of
the Dirac operator:
\begin{equation}
(i\slpartial_x-g\slA(x)-m)G_{_{F,R}}(x,y)=i\delta^{(4)}(x-y)\; ,
\end{equation}
but with different boundary conditions. For instance, the boundary
condition for the retarded propagator is simply:
\begin{equation}
\lim_{x_0\to y_0^+} G_{_{R}}(x,y)={\boldsymbol\delta}(\x-\y)\gamma^0\; .
\end{equation}
The boundary conditions for the Feynman propagator are awkward and not
very illuminating.

From these propagators, it is customary to first perform a Fourier
transform:
\begin{equation}
G_{_{F,R}}(q,p)\equiv \int d^4x d^4y\; e^{iq\cdot x}e^{-ip\cdot y}
G_{_{F,R}}(x,y)\; ,
\end{equation}
and then to extract the ``scattering matrix'' ${\cal T}_{{F,R}}$ via:
\begin{eqnarray}
G_{_{F,R}}(q,p)=(2\pi)^4\delta(q-p)G_{_{F,R}}^0(p)
+G_{_{F,R}}^0(q){\cal T}_{_{F,R}}(q,p)G_{_{F,R}}^0(p)\; ,
\label{eq:T-def}
\end{eqnarray}
where $G_{_{F,R}}^0$ is the {\sl free} Feynman or retarded quark
propagator. This definition removes the free term, and amputates the
external legs of the propagator.

Several quantities related to pair production in the external
classical field can then be expressed simply in terms of these
scattering matrices. The probability for producing a single $q\bar{q}$
pair is given by\footnote{From eqs.~(\ref{eq:T-def}) and
(\ref{eq:P1-def}), and given the fact that the canonical dimension of
the spinors $\overline{u}(\q)$ and $v(\p)$ is $({\rm momentum})^{1/2}$,
one can check that $P_1$ is indeed dimensionless.}:
\begin{equation}
P_1\left[A^\mu\right]=\left|\left<0_{\rm in}|0_{\rm out}\right>\right|^2
\int \frac{d^3\p}{(2\pi)^3 2\omega_\p}
\int \frac{d^3\q}{(2\pi)^3 2\omega_\q}
\left|
\overline{u}(\q){\cal T}_{_{F}}(q,-p)v(\p)
\right|^2\; ,
\label{eq:P1-def}
\end{equation}
where the argument $A^\mu$ has been used to remind the reader that
this is a quantity defined for one particular configuration of the
external classical field. The field $A^\mu$ is a functional of the
hard sources $\rho_1$ and $\rho_2$ (one for each nucleus) that
generate this classical field. The prefactor\footnote{This prefactor
is also the probability that no pair is produced during the collision,
which explains why it has to be smaller than $1$. Diagrammatically,
this quantity is the sum of the ``vacuum-vacuum'' diagrams, that have
no external legs besides those that connect to the external
field. More precisely, if we denote $V$ the sum of all {\sl connected}
vacuum-vacuum diagrams, then $|\langle 0_{\rm in}|0_{\rm
out}\rangle|^2=\exp(-2{\rm
Im}\,V)$~\cite{BaltzGMP1,GelisP1,GelisP2}. The expansion of ${\rm
Im}\,V$ in powers of the hard color sources starts at the order ${\cal
O}(\rho_1^2\rho_2^2)$.}  $\left|\left<0_{\rm in}|0_{\rm
out}\right>\right|^2$ is the square of the overlap between the vacua
at $x_0=-\infty$ and $x_0=+\infty$. This quantity is strictly smaller
than unity in an external field that can produce pairs. It is
necessary for unitarity to be preserved (for instance to ensure that
the {\sl probability} $P_1\left[A^\mu\right]$ is always smaller than
unity).  The cross-section for single pair production can be obtained
from here by:
\begin{equation}
\sigma_1=\int d^2\b \int [{\cal D}\rho_1{\cal D}\rho_2]\, 
W[\rho_1,\rho_2;\b]\, P_1\left[A^\mu[\rho_1,\rho_2]\right]\; ,
\label{eq:sigma1-def}
\end{equation}
where $W[\rho_1,\rho_2;\b]$ is the functional weight that defines the
statistical distribution of the hard sources at a given energy. This
functional depends on the impact parameter $\b$ between the two
nuclei. The latter must be integrated out in order to convert the
probability $P_1\left[A^\mu\right]$ into a cross-section.

Similarly, the average number of produced pairs in a given external
field configuration is given by:
\begin{equation}
\overline{n}\left[A^\mu\right]=
\int \frac{d^3\p}{(2\pi)^3 2\omega_\p}
\int \frac{d^3\q}{(2\pi)^3 2\omega_\q}
\left|
\overline{u}(\q){\cal T}_{_{R}}(q,-p)v(\p)
\right|^2\; .
\label{eq:nbar-def}
\end{equation}
The average number of pairs produced in a collision at a given impact
parameter $\b$ is given by:
\begin{equation}
\overline{n}(\b)=\int [{\cal D}\rho_1{\cal D}\rho_2] \,
W[\rho_1,\rho_2;\b]\, \overline{n}\left[A^\mu[\rho_1,\rho_2]\right]\; .
\end{equation}
Note also that both $P_1$ and $\overline{n}$ can be expressed as
differential distributions by undoing the integrations over the
momenta $\q$ and $\p$ of the quark and/or the antiquark.

When we give explicit expressions for the average over the
distribution of the hard color source charge densities, $\rho_1$ and $\rho_2$, we use
the McLerran-Venugopalan Gaussian model for the functional
$W[\rho_1,\rho_2;\b]$, which reads:
\begin{equation}
W[\rho_1,\rho_2;\b]=\exp
-\int d^2\x_\perp \left[
\frac{\rho_{1,a}(\x_\perp)\rho_{1,a}(\x_\perp)}{2\mu_1^2(\x_\perp)}+
\frac{\rho_{2,a}(\x_\perp)\rho_{2,a}(\x_\perp)}{2\mu_2^2(\x_\perp-\b)}
\right]\; ,
\label{eq:W-MV-def}
\end{equation}
where the functions $\mu_1^2(\x_\perp)$ and $\mu_2^2(\x_\perp)$ correspond to 
the color charges squared per unit area of the two nuclei. They are 
functions of the transverse coordinate that describe the number of
hard color sources per unit area at the transverse coordinate
$\x_\perp$. (The origin of the transverse coordinates is taken at the
center of the first nucleus, hence the argument $\x_\perp-\b$ for the
function that describes the second nucleus.) The canonical dimension
of $\mu_{1,2}^2$ is $({\rm momentum})^2$. In a model of a nucleus with
sharp edges, these functions would be almost flat inside the nuclear
disc and zero outside.

In the MV model, $\mu^2$ is energy (or x) independent. The $x$
dependence comes in through quantum
evolution~\cite{JalilKLW1,JalilKLW2,JalilKLW3,JalilKLW4,KovneM1,KovneMW3,Balit1,Kovch3,IancuLM1,IancuLM2}. In
the MV model, one can relate $\mu^2$ to the saturation scale
$Q_s$~\footnote{For a discussion of different conventions for $\mu^2$,
see Ref.~\cite{KrasnNV1}.}: \be Q_s^2(x,x_\perp) = \alpha_s N_c
\mu^2(x,x_\perp) \ln\left({g^2\mu^2\over \Lambda_{QCD}^2}\right) \, .
\ee The relation between the two scales is more non-trivial under
quantum evolution since the color screening scale is itself of order
$Q_s$~\cite{IancuIM1,IancuIM2,MuellT1,Trian1,IancuM2}.  We will see
later in Section 3 that, in our formalism, the $x$ dependence of
pair-production enters only through the $x$ dependence of the
saturation scale.

\subsection{Leading order approximation}
The relations of eqs.~(\ref{eq:P1-def}) and (\ref{eq:nbar-def}) are
completely general and are valid to all orders in the sources $\rho_1$
and $\rho_2$. In QCD, the external field $A^\mu[\rho_1,\rho_2]$ is a
nonlinear functional of the sources $\rho_1$ and $\rho_2$, that
receives contributions to all orders in $\rho_1$ and $\rho_2$. In
addition, the scattering matrices ${\cal T}_{_{F,R}}$ are themselves
functionals of the external field $A^\mu$, receiving contributions to
all orders in $A^\mu$. These facts make the quantities
$P_1\left[A^\mu\right]$ and $\overline{n}\left[A^\mu\right]$ extremely
complicated functionals of $\rho_1$ and $\rho_2$ that cannot be
calculated analytically. A complete solution would likely require a
numerical solution~\cite{GelisKL1}.

The leading term in the hard sources can however be calculated in
closed form. This approximation is justified if there is a scale in
the problem (the quark mass or the quark transverse momentum) which is
large compared to the scale set by the density of hard color sources
(namely, the saturation momentum $Q_s$).  We will assume that the
strong coupling constant $\alpha_s$ is sufficiently small to justify
keeping only the lowest order in $\alpha_s$. However, all corrections
of order $\alpha_s\ln(s)$ are included since they are resummed via the
quantum evolution of the functional $W[\rho_1,\rho_2;\b]$.

One can convince oneself that both $P_1\left[A^\mu\right]$ and
$\overline{n}\left[A^\mu\right]$ start at the order ${\cal
  O}(\rho_1^2\rho_2^2)$ in the hard sources. This is so because these
quantities are squares, and because the quark line must be attached at
least once to each nucleus if the pair is to be produced on-shell.
At this order in the hard color sources, there are several important
simplifications that are worth mentioning. Firstly, the prefactor
$\left|\left<0_{\rm in}|0_{\rm out}\right>\right|^2$ that appears in
$P_1$ can be replaced by $1$. Indeed, one has $\left|\left<0_{\rm
in}|0_{\rm out}\right>\right|^2=1+{\cal O}(\rho_1^2\rho_2^2)$.
Keeping higher order corrections in this prefactor would only produce
terms of order higher than ${\cal O}(\rho_1^2\rho_2^2)$ in $P_1$. The
second simplification occurs because there is no difference between the
time-ordered and retarded amplitudes. Indeed, since the probabilities
$P_2,P_3,\cdots$ to produce two or more pairs in a collision are of
higher order, we find 
\begin{eqnarray}
\overline{n}\left[A^\mu\right]&\equiv& P_1\left[A^\mu\right]+2P_2\left[A^\mu\right]+3P_3\left[A^\mu\right]+\cdots\nonumber\\
&\approx&P_1\left[A^\mu\right]\; .
\end{eqnarray}
The time-ordered and retarded amplitudes are the same at this order
because, in the language of Feynman diagrams, none of the intermediate
state quark propagators can be on their mass-shell (namely, the
$i\epsilon$ prescription of the quark propagators does not matter).

\begin{figure}[htbp]
\begin{center}
\resizebox*{!}{2.54cm}{\includegraphics{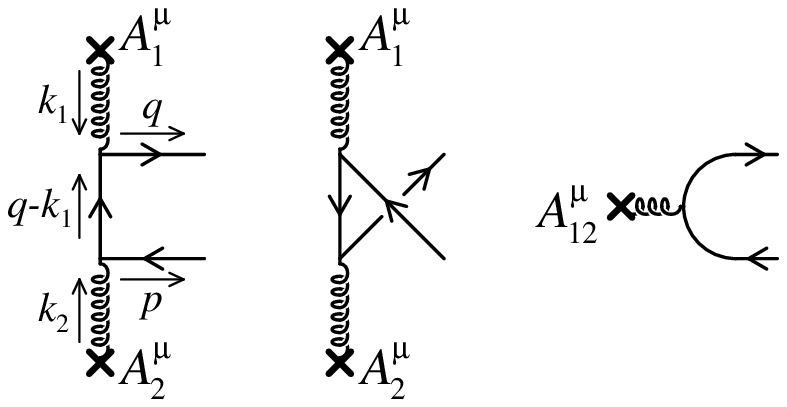}}
\end{center}
\caption{\label{fig:classical-diagrams} The leading contributions to
the pair production amplitude in terms of $A^\mu_1$, $A^\mu_2$ and
$A^\mu_{12}$. The gluon line terminated by a cross represents an
insertion $-igA^\mu(x)$ of the external field.}
\end{figure}
Our first task is therefore to obtain the classical field
$A^\mu[\rho_1,\rho_2]$ up to the order ${\cal O}(\rho_1\rho_2)$. We can
formally write:
\begin{equation}
A^\mu(x)\equiv A^\mu_1(x)+A^\mu_2(x)+A^\mu_{12}(x)+\cdots\; ,
\end{equation}
where $A^\mu_1(x)$ is the contribution of order ${\cal O}(\rho_1)$ to
the classical field, $A^\mu_2(x)$ the term of order ${\cal
O}(\rho_2)$, and $A^\mu_{12}(x)$ the term of order ${\cal
O}(\rho_1\rho_2)$. In terms of these contributions to the classical
field, we have to evaluate the three terms in figure
\ref{fig:classical-diagrams}. The derivation of the classical color
field in the covariant gauge up to the order ${\cal O}(\rho_1\rho_2)$
can be found in \cite{KovchR1}. We merely quote the results in momentum
space here, and refer the reader to the original literature or to
appendix \ref{sec:A} of the present paper where this derivation has
been reproduced in our notation\footnote{The light-cone coordinates
are defined as $x^{\pm}\equiv (x^0\pm x^3)/\sqrt{2}$. With this
convention, the invariant scalar product of two 4-vectors is $a\cdot
b=a^+b^-+a^-b^+-\a_\perp\cdot\b_\perp$ and the element of 4-volume is
$d^4x=dx^+dx^-d^2\x_\perp$.}:
\begin{eqnarray}
&&
A^+_{1,a}(k)=2\pi g\delta(k^-)\frac{1}{\k_\perp^2}\rho_{1,a}(\k_\perp)\; ,
\quad A^-_{1,a}(k)=A^i_{1,a}(k)=0\; ,
\nonumber\\
&&A^-_{2,a}(k)=2\pi g\delta(k^+)\frac{1}{\k_\perp^2}\rho_{2,a}(\k_\perp)\; ,
\quad A^+_{2,a}(k)=A^i_{2,a}(k)=0\; ,
\label{eq:A1A2}
\end{eqnarray}
\begin{eqnarray}
&&A^+_{12,a}(k)=-ig\frac{f^{abc}}{k^2}\int\frac{d^4k_1}{(2\pi)^4}
\left\{k_1^++\frac{k_1^2}{k_2^-}
\right\}A^+_{1,b}(k_1)A^-_{2,c}(k_2)\; ,\nonumber\\
&&A^-_{12,a}(k)=ig\frac{f^{abc}}{k^2}\int\frac{d^4k_1}{(2\pi)^4}
\left\{k_2^-+\frac{k_2^2}{k_1^+}
\right\}A^+_{1,b}(k_1)A^-_{2,c}(k_2)\; ,\nonumber\\
&&A^i_{12,a}(k)=ig\frac{f^{abc}}{k^2}\int\frac{d^4k_1}{(2\pi)^4}
\left\{k_2^i-k_1^i
\right\}A^+_{1,b}(k_1)A^-_{2,c}(k_2)\; ,
\label{eq:A12}
\end{eqnarray}
where $k_2\equiv k-k_1$.

\subsection{Pair production amplitude}
Following figure \ref{fig:classical-diagrams}, we can write the
leading order pair production amplitude as:
\begin{equation}
\overline{u}(\q){\cal T}_{_{F,R}}v(\p)\equiv
{\cal M}(\q,\p)={\cal M}_{1+2}(\q,\p)+{\cal M}_{12}(\q,\p)\; ,
\end{equation}
where we have defined
\begin{eqnarray}
&&{\cal M}_{1+2}(\q,\p)\equiv
\int\frac{d^4k_1}{(2\pi)^4}(-ig A^\mu_{1,a}(k_1))(-ig A^\nu_{2,b}(k_2))
\nonumber\\
&&\qquad\qquad\times
\overline{u}(\q)\left\{\gamma_\mu t_a G^0(q-k_1)\gamma_\nu t_b
+
\gamma_\nu t_b G^0(k_1-p)\gamma_\mu t_a \right\} v(\p)
\end{eqnarray}
and
\begin{eqnarray}
{\cal M}_{12}(\q,\p)\equiv(-ig A^\mu_{12,a}(p+q)) \,
\overline{u}(\q)\gamma_\mu t_a v(\p)\; ,
\end{eqnarray}
where the $t_{a,b}$ are color matrices in the fundamental
representation. $G^0(k)$ is the free quark propagator\footnote{The
  prefactor $i$ in this propagator is purely conventional. However, it
  cannot be chosen independently of the $-i$ in the field insertion
  $-igA^\mu$.}:
\begin{equation}
G^0(k)\equiv i\frac{\slk+m}{k^2-m^2+i\epsilon}\; ,
\end{equation}
and $k_2\equiv p+q-k_1$. 

Using the fact that the momenta $p$ and $q$ are on-shell, that all but
one component of $A^\mu_1$ and $A^\mu_2$ are zero, and finally
$k_1^-=k_2^+=0$, we can obtain easily:
\begin{eqnarray}
&&{\cal M}_{1+2}(\q,\p)\equiv
ig^2\int\frac{d^4k_1}{(2\pi)^4}A^+_{1,a}(k_1)A^-_{2,b}(k_2)\nonumber\\
&&\qquad\qquad\qquad\quad\times
\overline{u}(\q)\left\{
\gamma^- 
\frac{m-{\boldsymbol\gamma}_\perp\cdot({\q_\perp-\k_{1\perp}})}
{2q^-p^++(\q_\perp-\k_{1\perp})^2+m^2}\gamma^+ t_a t_b\right.
\nonumber\\
&&
\qquad\qquad\qquad\qquad\qquad\left.+
\gamma^+
\frac{m+{\boldsymbol\gamma}_\perp\cdot({\p_\perp-\k_{1\perp}})}
{2q^+p^-+(\p_\perp-\k_{1\perp})^2+m^2}\gamma^- t_b t_a
\right\}v(\p)\; .\nonumber\\
&&
\end{eqnarray}
We readily see that the denominators in this expression are
positive definite, which justifies a posteriori the fact that the
$i\epsilon$ prescription in the free quark propagator is irrelevant at
this order. Using eqs.~(\ref{eq:A12}), we can similarly write the term
${\cal M}_{12}(\q,\p)$ as:
\begin{eqnarray}
&&{\cal M}_{12}(\q,\p)=ig^2\frac{\left[t_a,t_b\right]}{(p+q)^2}
\int\frac{d^4k_1}{(2\pi)^4}A^+_{1,a}(k_1)A^-_{2,b}(k_2)\nonumber\\
&&\quad\times
\overline{u}(\q)\Bigg\{
\gamma^-\left(k_1^+-\frac{\k_{1\perp}^2}{k_2^-}\right)
-\gamma^+\left(k_2^--\frac{\k_{2\perp}^2}{k_1^+}\right)
+{\boldsymbol\gamma}_\perp\cdot(\k_{1\perp}-\k_{2\perp})
\Bigg\}v(\p)\; .\nonumber\\
&&
\end{eqnarray}
Here again, all the denominators are strictly positive, making the
$i\epsilon$ prescriptions irrelevant. Note that we can write the
expression of ${\cal M}_{12}$ as follows:
\begin{eqnarray}
{\cal M}_{12}(\q,\p)=ig^2\frac{\left[t_a,t_b\right]}{(p+q)^2}
\int\frac{d^4k_1}{(2\pi)^4}A^+_{1,a}(k_1)A^-_{2,b}(k_2)
\;\overline{u}(\q)\, \slc \, v(\p)\; ,
\end{eqnarray}
where $c^\mu$ is the following 4-vector:
\begin{equation}
c\equiv\left(c^+\!=\!p^+\!+\!q^+\!-\!\frac{\k_{1\perp}^2}{p^-\!+\!q^-},
c^-\!=\!\frac{\k_{2\perp}^2}{p^+\!+\!q^+}\!-\!p^-\!-\!q^-,
\cc_\perp\!=\!\k_{2\perp}\!-\!\k_{1\perp}
\right)\; .
\end{equation}
We note that the vector $c^\mu$ is indeed nothing else but the standard effective
Lipatov vertex (see eq.~(80) of \cite{Duca1} for instance). In particular, one may
check that it obeys the transversality relation:
\begin{equation}
(p+q)\cdot c=0\; .
\end{equation}

At this stage, it is trivial to replace $A^+_1$ and $A^-_2$ by their
expression in terms of the source densities (eqs.~(\ref{eq:A1A2})) in
order to obtain the leading order pair production amplitude as a
functional of the hard color sources $\rho_1$ and $\rho_2$.

\subsection{Ward identity verification}
\label{sec:ward}
One can write the pair production amplitude ${\cal M}(\q,\p)$ as follows:
\begin{eqnarray}
{\cal M}(\q,\p)=\int\frac{d^4k_1}{(2\pi)^4}
A^\mu_{1,a}(k_1)A^\nu_{2,b}(k_2) m_{\mu\nu}^{ab}(k_1,k_2;\q,\p)\; .
\label{eq:mab-def}
\end{eqnarray}
The expressions of the previous subsection tell us that:
\begin{eqnarray}
&&m^{-+}_{ab}(k_1,k_2;\q,\p)=ig^2\nonumber\\
&&\!\!\times\overline{u}(\q)\Bigg\{
\frac{\gamma^-
(m\!-\!{\boldsymbol\gamma}_\perp\cdot({\q_\perp-\k_{1\perp}}))
\gamma^+ t_a t_b}
{2q^-p^++(\q_\perp-\k_{1\perp})^2+m^2}
+
\frac{
\gamma^+
(m\!+\!{\boldsymbol\gamma}_\perp\cdot({\p_\perp-\k_{1\perp}}))
\gamma^- t_b t_a
}
{2q^+p^-+(\p_\perp-\k_{1\perp})^2+m^2}
\nonumber\\
&&
\quad
+\frac{\left[t_a,t_b\right]}{(p+q)^2}\left[
\gamma^-\!\left[{k_1^+\!-\!\frac{\k_{1\perp}^2}{k_2^-}}\right]
\!-\!\gamma^+\!\left[{k_2^-\!-\!\frac{\k_{2\perp}^2}{k_1^+}}\right]
\!+\!{{\boldsymbol\gamma}_\perp\cdot(\k_{1\perp}\!-\!\k_{2\perp})}
\right]
\Bigg\}v(\p)\; ,\nonumber\\
&&
\end{eqnarray}
while the other components of $m^{\mu\nu}_{ab}$ do not appear in the
expression of the pair production amplitude. One notes that
$m^{\mu\nu}_{ab}(k_1,k_2;\q,\p)$ is {\sl not} the amplitude for the
process $gg\to q\bar{q}$. Indeed, contracting the $gg\to q\bar{q}$
amplitude with the classical fields $A^\mu_1(k_1)$ and $A^\nu_2(k_2)$
would only give the first three diagrams of figure
\ref{fig:feynman-diagrams}, and leave out the bremsstrahlung diagrams.

It is interesting to investigate the limit $\k_{1\perp}=
{\boldsymbol 0}$ of $m^{-+}_{ab}(k_1,k_2;\q,\p)$. We have:
\begin{eqnarray}
&&\lim_{\k_{1\perp}\to 0}m^{-+}_{ab}(k_1,k_2;\q,\p)=\nonumber\\
&&\quad=ig^2
\overline{u}(\q)\Bigg\{
\frac{\gamma^- (m-{\boldsymbol\gamma}_\perp\cdot{\q_\perp})\gamma^+ t_a t_b}
{2q^-p^++\q_\perp^2+m^2}+
\frac{\gamma^+(m+{\boldsymbol\gamma}_\perp\cdot{\p_\perp})\gamma^- t_b t_a}
{2q^+p^-+\p_\perp^2+m^2}
\nonumber\\
&&
\qquad\quad
+\frac{\left[t_a,t_b\right]}{(p+q)^2}\left[
\gamma^-{k_1^+}
-\gamma^+\left[{k_2^--\frac{\k_{2\perp}^2}{k_1^+}}\right]
-{{\boldsymbol\gamma}_\perp\cdot\k_{2\perp}}\right]
\Bigg\}v(\p)\; .
\end{eqnarray}
Using the Dirac equations obeyed by the spinors $\overline{u}(\q)$ and
$v(\p)$, as well as the following relations:
\begin{eqnarray}
&&\q_\perp^2+m^2=2q^+q^-\; ,\quad
\p_\perp^2+m^2=2p^+p^-\; ,\nonumber\\
&&k_1^+=p^++q^+\; ,\quad
k_2^-=p^-+q^-\; ,\quad\k_{2\perp}=\p_\perp+\q_\perp\;({\rm if\ }\k_{1\perp}=0)\nonumber\\
&&\gamma^+\gamma^-\gamma^+=2\gamma^+\; ,
\end{eqnarray}
we can simplify the above limit into:
\begin{eqnarray}
&&\lim_{\k_{1\perp}\to 0}m^{-+}_{ab}(k_1,k_2;\q,\p)=\nonumber\\
&&\quad=ig^2
\overline{u}(\q)\gamma^+ v(\p)
\frac{\left[t_a,t_b\right]}{p^++q^+}
\left\{1
+\left[
\frac{(\p_\perp+\q_\perp)^2}{p^++q^+}
-2(p^-+q^-)
\right]\frac{p^++q^+}{(p+q)^2}\right\}
\nonumber\\
&&\quad=0\; .
\end{eqnarray}
Similarly, one can check that:
\begin{eqnarray}
\lim_{\k_{2\perp}\to 0}m^{-+}_{ab}(k_1,k_2;\q,\p)=0\; .
\end{eqnarray}
These simple limits are consequences of the following Ward
identities\footnote{Because of eq.~(\ref{eq:mab-def}), the object
  $m_{ab}^{\mu\nu}(k_1,k_2;\q,\p)$ may be seen as the lowest order
  value of the correlator $\langle J^\mu_1 J^\nu_2
  \overline{\psi}\psi\rangle$ between the currents of the individual
  nuclei and a pair of fermions. The currents $J^\mu_1$ and $J^\nu_2$
  behave like Abelian currents at the order we are considering (the
  non Abelian term in the current conservation law is of higher order
  in the sources $\rho_{1,2}$), which is the reason why we have
  Abelian-like Ward identities for $m_{ab}^{\mu\nu}(k_1,k_2;\q,\p)$.}:
\begin{eqnarray}
k_{1\mu}\, m_{ab}^{\mu\nu}(k_1,k_2;\q,\p)
=k_{2\nu}\, m_{ab}^{\mu\nu}(k_1,k_2;\q,\p)=0\; .
\end{eqnarray}
Indeed, since $k_1^-=0$, the first one implies for instance:
\begin{eqnarray}
k_1^+\, m_{ab}^{-+}(k_1,k_2;\q,\p) =k_1^i\, m^{i
+}_{ab}(k_1,k_2;\q,\p)\empile{=}\over{\k_{1\perp}\to 0}{\cal O}(k_1^i)
\; .
\label{eq:ward}
\end{eqnarray}
In other words, $m_{ab}^{-+}(k_1,k_2;\q,\p)$ must vanish linearly with
$\k_{1\perp}$ when $\k_{1\perp}\to 0$ in order to fulfill the Ward
identity.  The physical meaning of these vanishing limits is that in
order to produce the quark-antiquark pair on-shell, some transverse
momentum has to come from both nuclei. As we shall see later, this
property is also essential in order to soften some collinear
singularities.

\section{Pair production probability}
\subsection{Average over the hard color sources}
From eq.~(\ref{eq:mab-def}), we can write the pair production
amplitude in terms of the hard color sources:
\begin{eqnarray}
{\cal M}(\q,\p)=g^2\int\frac{d^2\k_{1\perp}}{(2\pi)^2} 
\frac{\rho_{1,a}(\k_{1\perp})}{\k_{1\perp}^2}
\frac{\rho_{2,b}(\k_{2\perp})}{\k_{2\perp}^2}
m^{-+}_{ab}(k_1,k_2;\q,\p)\; .
\end{eqnarray}
Squaring this amplitude, and averaging over the sources, we obtain:
\begin{eqnarray}
&&P_1(\b)=\overline{n}(\b)=
g^4\int \frac{d^3\p}{(2\pi)^3 2\omega_\p}
\int \frac{d^3\q}{(2\pi)^3 2\omega_\q}
\int\left[{\cal D}\rho_1{\cal D}\rho_2\right]\;
W\left[\rho_1,\rho_2;\b\right]\nonumber\\
&&\qquad\qquad\qquad\times
\int\frac{d^2\k_{1\perp}}{(2\pi)^2}
\frac{d^2\k^\prime_{1\perp}}{(2\pi)^2}
\frac{\rho_{1,a}(\k_{1\perp})}{\k_{1\perp}^2}
\frac{\rho_{2,b}(\k_{2\perp})}{\k_{2\perp}^2}
\frac{\rho_{1,a^\prime}^*(\k^\prime_{1\perp})}{\k_{1\perp}^{\prime 2}}
\frac{\rho_{2,b^\prime}^*(\k^\prime_{2\perp})}{\k_{2\perp}^{\prime 2}}
\nonumber\\
&&\qquad\qquad\qquad\qquad\times
{\rm Tr}\,\left(m^{-+}_{ab}(k_1,k_2;\q,\p)
m^{-+\ *}_{a^\prime b^\prime}(k^\prime_1,k^\prime_2;\q,\p)\right)\; ,
\end{eqnarray}
where ${\rm Tr}$ denotes a trace of the color and Dirac matrices.  In
the McLerran-Venugopalan model where the functional
$W[\rho_1,\rho_2;\b]$ is given by eq.~(\ref{eq:W-MV-def}), the source
averages are given in coordinate space by:
\begin{eqnarray}
&&\int \left[{\cal D}\rho_1{\cal D}\rho_2\right]\;
W\left[\rho_1,\rho_2;\b\right]
\rho_{1,a}(\x_\perp)\rho_{1,a^\prime}(\x^\prime_\perp)
\rho_{2,b}(\y_\perp)\rho_{2,b^\prime}(\y^\prime_\perp)
=\nonumber\\
&&\qquad\qquad=\delta^{aa^\prime}\delta^{bb^\prime}
\mu^2_1(\x_\perp)
\mu^2_2(\y_\perp-\b)
\delta(\x_\perp-\x^\prime_\perp)\delta(\y_\perp-\y^\prime_\perp)\; .
\label{eq:color-avg}
\end{eqnarray}
The Fourier transform of eq.~(\ref{eq:color-avg}) reads:
\begin{eqnarray}
&&\int \left[{\cal D}\rho_1{\cal D}\rho_2\right]\;
W\left[\rho_1,\rho_2;\b\right]
\rho_{1,a}(\k_{1\perp})
\rho_{2,b}(\k_{2\perp})
\rho_{1,a^\prime}^*(\k^\prime_{1\perp})
\rho_{2,b^\prime}^*(\k^\prime_{2\perp})=\nonumber\\
&&\qquad\qquad=\delta^{aa^\prime}\delta^{bb^\prime}
e^{-i\b\cdot(\k_{1\perp}-\k_{1\perp}^\prime)}
\widetilde{\mu_1^2}(\k_{1\perp}-\k^\prime_{1\perp})
\widetilde{\mu_2^2}(\k^\prime_{1\perp}-\k_{1\perp})\; ,
\end{eqnarray}
where $\widetilde{\mu_{1,2}^2}(\k_\perp)$ is the Fourier transform of
$\mu_{1,2}^2(\x_\perp)$. We have used the fact that
$\k_{2\perp}=\p_\perp+\q_\perp-\k_{1\perp}$. 

\subsection{Impact parameter and energy dependence}
At this stage, we have the following expression for the pair
production probability:
\begin{eqnarray}
&&P_1(\b)=\overline{n}(\b)=
g^4\int \frac{d^3\p}{(2\pi)^3 2\omega_\p}
\int \frac{d^3\q}{(2\pi)^3 2\omega_\q}
\int\frac{d^2\k_{1\perp}}{(2\pi)^2}
\frac{d^2\k^\prime_{1\perp}}{(2\pi)^2}
e^{-i\b\cdot(\k_{1\perp}-\k_{1\perp}^\prime)}
\nonumber\\
&&\times
\widetilde{\mu_1^2}(\k_{1\perp}-\k^\prime_{1\perp})
\widetilde{\mu_2^2}(\k^\prime_{1\perp}-\k_{1\perp})
\frac{{\rm Tr}\,\left(m^{-+}_{ab}(k_1,k_2;\q,\p)
m^{-+\ *}_{ab}(k^\prime_1,k^\prime_2;\q,\p)\right)}
{\k_{1\perp}^2\k^{\prime 2}_{1\perp}\k^2_{2\perp}\k^{\prime 2}_{2\perp}}
\; .\nonumber\\
&&
\end{eqnarray}
For large nuclei, these Fourier transforms are strongly peaked around
$\k_\perp=0$, with a typical width of the order of $1/R$ where $R$ is
the radius of the nucleus. For instance, for $R=6{\rm\ fm}$, the
spread in transverse momentum would be of order $\Delta\k_\perp\sim
1/R \sim 30{\rm \ MeV}$. The factors $m_{ab}^{-+}$ contain a large
momentum scale provided by the mass $m$ of the heavy quarks, which we
assume to be much larger than $1/R$. Therefore, we can safely neglect
the difference $\k_{1\perp}-\k^\prime_{1\perp}$ in these
factors. Regarding the denominators $\k_{1\perp}^2\k^{\prime
2}_{1\perp}\k^2_{2\perp}\k^{\prime 2}_{2\perp}$, the situation is more
delicate. Indeed, they could be arbitrarily small because of the
presence of a collinear singularity when $\k_{1\perp}\to 0$ or
$\k_{1\perp}\to \p_\perp+\q_\perp$. We are going to assume that these
collinear singularities are regularised by physics at the scale of the
saturation momentum $Q_s\gg 1/R$, so that we can also neglect the
difference $\k_{1\perp}-\k^\prime_{1\perp}$ in the denominators.
Therefore, we can easily integrate out the difference
$\k_{1\perp}-\k^\prime_{1\perp}$ in order to get the impact parameter
dependence of the pair production probability:
\begin{eqnarray}
&&P_1(\b)=\overline{n}(\b)=
g^4\left[\int d^2\x_\perp\mu_1^2(\x_\perp)\mu_2^2(\x_\perp-\b)\right]
\nonumber\\
&&\quad\times
\int \frac{d^3\p}{(2\pi)^3 2\omega_\p}
\int \frac{d^3\q}{(2\pi)^3 2\omega_\q}
\int\frac{d^2\k_{1\perp}}{(2\pi)^2}
\frac{{\rm Tr}\,\left(\left|m^{-+}_{ab}(k_1,k_2;\q,\p)\right|^2\right)}
{\k_{1\perp}^4\k_{2\perp}^4}
\; .
\label{eq:P1-1}
\end{eqnarray}
The prefactor between the square brackets is the overlap of the
density functions that describe the two colliding nuclei. In a crude
model of ``cylindrical'' nuclei, it would be equal to $\mu_1^2\mu_2^2
S(\b)$ where $S(\b)$ is the area of the overlap between the two
nuclei. The explicit expression of the trace that appears in this
formula is given in the appendix \ref{sec:traces}.

Since there is no quantum evolution in the MV model, the functions
$\mu_1^2(\x_\perp)$ and $\mu_2^2(\x_\perp)$ do not depend on the
longitudinal momentum fraction of the gluons. As a consequence, the
differential pair production probability is boost invariant, in the
sense that it depends only on the rapidity difference $y_p-y_q$ of the
quark and the antiquark. Strictly speaking, eq.~(\ref{eq:P1-1}) gives
an infinite result because it contains also an integration over the
mean rapidity $(y_p+y_q)/2$: this infinity should be cutoff by the
fact that the rapidity of the produced quark and antiquark are limited
by the rapidity of the projectiles. More formally, quantum evolution
of the functional $W[\rho_1,\rho_2;\b]$ would introduce a dependence
of the functions $\mu_1^2(\x_\perp)$ and $\mu_2^2(\x_\perp)$ on the
gluon momentum fractions, such that they vanish when we reach the
fragmentation region of the corresponding projectile.

\section{Relation to $k_\perp$-factorization and Collinear factorization}
\begin{figure}[htbp]
\begin{center}
\resizebox*{!}{2.2cm}{\includegraphics{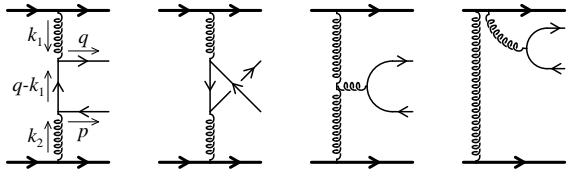}}
\end{center}
\caption{\label{fig:feynman-diagrams} The Feynman diagrams that can
contribute to pair production at the order ${\cal
O}(\rho_1\rho_2)$. The bold lines represent the classical sources
$\rho_1$ and $\rho_2$.}
\end{figure}

In $k_\perp$-factorized perturbation theory, it is well known that the
leading order diagrams contributing to the production of a
quark-antiquark pair are those listed in figure
\ref{fig:feynman-diagrams}.  The first two diagrams are the analogs of
the diagrams one would have in QED, except for the fact that the
coupling of the gluon to the quark line involves a non-commuting SU(3)
matrix. The third diagram involves the 3-gluon vertex, and the fourth
diagram\footnote{There are four bremsstrahlung diagrams of this type,
and only one has been represented.} is a bremsstrahlung contribution
required in order to ensure gauge invariance.

It is well known that the first two terms of figure
\ref{fig:classical-diagrams} correspond to the first two Feynman
diagrams of figure \ref{fig:feynman-diagrams} in the limit where the
transverse momentum of the two gluons is small in front of the
longitudinal momentum of the hard sources.  Moreover, it has been
shown by Kovchegov and Rischke \cite{KovchR1} that the term
proportional to $A^\mu_{12}(x)$ contains both the diagram with the
3-gluon vertex and the bremsstrahlung diagrams. Therefore, it is very
interesting to compare the result of our classical calculation to the
standard result obtained in the framework of $k_\perp$-factorization
\cite{ColliE1,CatanCH1}.

It turns out that the formula we have just derived has a fairly simple
connection with the result one would obtain in the framework of
$k_\perp$-factorization. In order to see this connection, one needs
the following relationship between the quantities
$\mu_{1,2}^2(\x_\perp)$ and the unintegrated gluon
distribution\footnote{There are many definitions of the unintegrated
gluon distribution in the literature. The one we are using here is:
\begin{equation}
xG(x,Q^2)\equiv \int_0^{Q^2}d(k_\perp^2) \varphi(x,k_\perp)\; .
\end{equation}
}
$\varphi_{1,2}(k_\perp)$:
\begin{equation}
\frac{d\varphi(k_\perp,\x_\perp)}{d^2\x_\perp}
=\pi g^2 d_{_{A}}\frac{\mu^2(\x_\perp)}{\k_\perp^2}\; ,
\end{equation}
where $d\varphi(k_\perp,\x_\perp)/d^2\x_\perp$ is the number of gluons
in a projectile, per unit of $k_\perp^2$ and per unit area, and where
$d_{_{A}}\equiv N_c^2-1$ is the dimension of the adjoint
representation. In fact, this identification of the unintegrated gluon
distribution is more general than the model of
eq.~(\ref{eq:color-avg}), for one has in general:
\begin{eqnarray}
\frac{d\varphi(k_\perp,\x_\perp)}{d^2\x_\perp}=
\pi\frac{1}{\k_\perp^2}\int d^2\r_\perp
e^{-i\k_\perp\cdot \r_\perp}
\left<\rho_a(\x_\perp+\r_\perp/2)\rho_a(\x_\perp-\r_\perp/2)\right>_\rho\; ,
\label{eq:rhos}
\end{eqnarray}
where $\langle\cdots\rangle_\rho$ denotes the average over the hard
color sources.  From there, using eqs.~(\ref{eq:sigma1-def}) and
(\ref{eq:P1-1}), it is easy to write the pair production cross-section
as:
\begin{eqnarray}
&&\frac{d\sigma_1}{dy_pdy_qd^2\p_\perp d^2\q_\perp}=
\frac{1}{(2\pi)^6 d_{_{A}}^2}
\int\frac{d^2\k_{1\perp}}{(2\pi)^2}
\frac{d^2\k_{2\perp}}{(2\pi)^2}
\delta(\k_{1\perp}+\k_{2\perp}-\p_\perp-\q_\perp)\nonumber\\
&&\times
\int d^2\b d^2\x_\perp 
\frac{d\varphi_1(k_{1\perp},\x_\perp)}{d^2\x_\perp}
\frac{d\varphi_2(k_{2\perp},\x_\perp-\b)}{d^2\x_\perp}
\frac{{\rm Tr}\,\left(\left|m^{-+}_{ab}(k_1,k_2;\q,\p)\right|^2\right)}
{\k_{1\perp}^2 \k_{2\perp}^2}
\; ,\nonumber\\
&&
\end{eqnarray}
where $y_q$ and $y_p$ are the rapidities of the quark and the
antiquark.  The integration over $\b$ and $\x_\perp$ is trivial to
perform, and leads to the usual unintegrated gluon distributions:
\begin{eqnarray}
&&\frac{d\sigma_1}{dy_pdy_qd^2\p_\perp d^2\q_\perp}
=\frac{1}{(2\pi)^6 d_{_{A}}^2}
\int\frac{d^2\k_{1\perp}}{(2\pi)^2}\frac{d^2\k_{2\perp}}{(2\pi)^2}
\delta(\k_{1\perp}+\k_{2\perp}-\p_\perp-\q_\perp)\nonumber\\
&&\qquad\qquad\times \varphi_1(k_{1\perp}) \varphi_2(k_{2\perp})
\frac{{\rm Tr}\,\left(\left|m^{-+}_{ab}(k_1,k_2;\q,\p)\right|^2\right)}
{\k_{1\perp}^2
\k_{2\perp}^2} \; .
\end{eqnarray}
This expression has a structure which is very similar to what one
would have in the $k_\perp$-factorization approach.  Moreover, we have
checked (see the appendix \ref{sec:traces}) that the matrix element
given by ${\rm
Tr}\,\big(\left|m^{-+}_{ab}(k_1,k_2;\q,\p)\right|^2\big)$ is exactly
identical to the matrix element obtained in the framework of
$k_\perp$-factorization by Collins and Ellis \cite{ColliE1}. Note also
that the $x$ dependence here comes in through the unintegrated gluon
distribution.  This $x$ dependence of unintegrated gluon distribution
is related to that of the correlator of $\rho$'s by
eq.~\ref{eq:rhos}. The correlator in turn is determined by the weight
function $W$ in eq.~\ref{eq:sigma1-def} which, as discussed
previously, satisfies a non-linear renormalization group equation in
$x$.

From the previous formula, it is well known how to recover the
standard results of collinear factorization. One must take the limit
$|\k_{1\perp}|,|\k_{2\perp}|\to 0$ in the quantity ${\rm
Tr}\,\big(\left|m^{-+}_{ab}(k_1,k_2;\q,\p)\right|^2\big)/\k_1^2\k_2^2$
(but not in the unintegrated gluon distributions). This limit is
perfectly defined thanks to the Ward identities discussed in section
\ref{sec:ward}. Then, the integration over the azimuthal angles of the
vectors $\k_{1\perp}$ and $\k_{2\perp}$ gives the expression of the
matrix element $gg\to q\bar{q}$ in the limit of collinear
factorization, while the integration of the non integrated gluon
distributions over $k_{1\perp}^2$ and $k_{2\perp}^2$ reconstructs the
integrated gluon distributions.  After this procedure, we are left
with a factor $\delta(\p_\perp+\q_\perp)$ which naturally corresponds
to the fact that in this limit the quark and antiquark must be
produced back-to-back in the transverse plane.

\section{Beyond leading order}
\begin{figure}[htbp]
\begin{center}
\resizebox*{!}{4.92cm}{\includegraphics{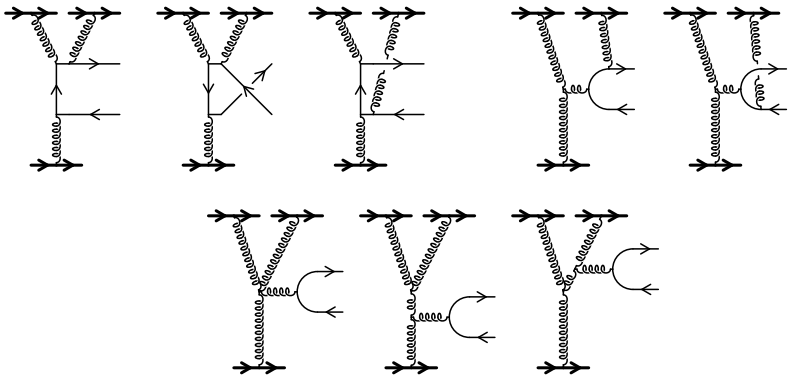}}
\end{center}
\caption{\label{fig:diagrams-NLO} Contributions to the pair production
amplitude at order ${\cal O}(\rho_1^2 \rho_2)$. The diagrams involving
bremsstrahlung have been omitted. The bold lines represent the hard
color sources.}
\end{figure}

Corrections due to terms of higher order in the hard color sources,
breaking the $k_\perp$ factorization, are expected to become important
in the soft regime, i.e. when the transverse mass of the quark or
antiquark is of the order of the saturation momentum or smaller. In
this section, we discuss the main issues that arise when one
calculates corrections to the pair production amplitude that are of
higher order in the hard color sources. In order to illustrate this
discussion, let us first consider the example of corrections of order
${\cal O}(\rho_1^2\rho_2)$ to the amplitude, which are represented in
figure \ref{fig:diagrams-NLO}.
The contributions at this order can be grouped in three sets,
according to the number of classical field insertions along the quark
line. The first set (the three diagrams in the upper left part of
figure \ref{fig:diagrams-NLO}) involves only the Weizs\"acker-Williams
fields $A_1^\mu$ and $A_2^\mu$ produced by the individual nuclei. The
only difference with the previous order is that there are now two
powers of $A_1^\mu$ and one power of $A_2^\mu$. Then, there are two
diagrams (represented in the upper right part of figure
\ref{fig:diagrams-NLO}) involving one power of the classical field
$A_{12}^\mu$ and one power of the field $A_1^\mu$. Finally, we have
three terms with a single insertion of the classical field
$A_{112}^\mu$, i.e. the term of order ${\cal O}(\rho_1^2\rho_2)$ in
the solution of the classical Yang-Mills equation.

Beyond leading order, there also arises the issue of the vacuum-vacuum
diagrams contained in the prefactor $\left|\left<0_{\rm in}|0_{\rm
      out}\right>\right|^2$, since this factor cannot be replaced by
$1$ anymore. Although this factor can be calculated order by order in
an expansion in powers of the hard color sources, this makes the
calculation of the single pair production probability much more
difficult than the calculation of the average pair multiplicity.
Alternatively, one could be less rigorous and simply discard this
prefactor in the calculation of $P_1$. This would of course violate
unitarity (thereby resulting in a probability $P_1$ larger than
unity), which could then be restored in an approximate way by the
replacement $P_1\to P_1\exp(-P_1)$. This way of restoring unitarity is
equivalent to assuming that the distribution of pair multiplicities is
a Poisson distribution.

With this in mind, in order to perform the calculation of pair
production beyond the leading order in the hard sources, one needs to
go through the following steps:
\begin{itemize}
\item solve the classical Yang-Mills equation with two sources up to
the required order. This can in principle be done analytically by a
generalization of the method presented in appendix \ref{sec:A}. But
for a non-perturbative calculation to all orders, one has to resort to
a numerical resolution of the classical Yang-Mills equation with
retarded boundary conditions. This step is under control after the
work performed in \cite{KrasnV1,KrasnV2,KrasnNV2,Lappi1}.
\item compute the propagator of a quark in the previously obtained
classical field. Only terms whose order is not higher than ${\cal
O}(\rho_1^2\rho_2)$ need to be kept in the above example. For a
systematic study of higher orders, one must rewrite the retarded
amplitude $\overline{u}(\q){\cal T}_{_{R}}v(\p)$ in terms of retarded
solutions of the Dirac equation, and then solve this equation
numerically \cite{GelisKL1}. Here it is crucial to limit ourselves to
the calculation of the pair multiplicity, which requires only the
retarded quark propagator in the external field. Indeed, the numerical
determination of the time-ordered propagator is a much harder problem,
due to the great complexity of the boundary conditions obeyed by this
type of propagator.
\item study the quantum evolution of the functional
  $W[\rho_1,\rho_2;\b]$.  Contrary to the leading order where only the
  correlators $\big<\rho_{1,2}\rho_{1,2}\big>$ were necessary, we need
  now higher order correlators. For instance, in the calculation at
  order ${\cal O}(\rho_1^2\rho_2)$ considered as an example, we would
  need the correlators $\big<\rho_1\rho_1\rho_1\big>$ ad
  $\big<\rho_1\rho_1\rho_1\rho_1\big>$. Indeed, strictly speaking
  quantum evolution does not preserve the Gaussian structure (unless
  in certain specific limits \cite{IancuIM2}). There are now ways to
  evaluate numerically these correlators \cite{RummuW1}.
  Alternatively, at collision energies that are not too large, this
  step can be skipped since a Gaussian weight is a good approximation.
\end{itemize}

\section{Conclusions}
We have discussed in this paper heavy quark pair production in heavy
ion collisions, in the framework of the Color Glass Condensate model.
We have computed explicitly the leading order contribution to the
heavy quark pair production cross-section.  This computation is valid
in the kinematic region where $k_\perp \gg Q_s$. We show explicitly
that the classical result in this approximation is identical to the
$k_\perp$ factorization results derived previously. We discuss how
$k_\perp$ factorization may be violated for smaller $k_\perp$ as one
enters the saturation regime. Numerical computations addressing this
issue are in progress and will be reported separately~\cite{GelisKL1}.

We have not attempted here to relate our results to phenomenology.
There have already been several phenomenological works applying
$k_\perp$ factorization to various aspects of heavy quark production
in Deeply Inelastic Scattering and in hadron-hadron scattering
experiments.  $k_\perp$ factorization ideas are also being studied in
the framework of relativistic heavy ion collisions~\footnote{Private
  communication from D. Kharzeev and K. Tuchin.}.

\section*{Acknowledgements}
We thank K. Kajantie, D. Kharzeev, T. Lappi, E. M. Levin and K. Tuchin
for useful discussions.  R.~V.'s research was supported by DOE
Contract No. DE-AC02-98CH10886 in part by the RIKEN-BNL Research
Center.

\appendix

\section{Classical color field to order ${\cal O}(\rho_1\rho_2)$}
\label{sec:A}
We derive the classical field in the covariant gauge, and we follow
closely the method of Kovchegov and Rischke. The gauge field is a
solution of the classical Yang-Mills equation:
\begin{equation}
\left[D_\mu,F^{\mu\nu}\right]=J^\nu\; ,
\label{eq:YM}
\end{equation}
where $J^\nu$ is the classical current which must be covariantly
conserved:
\begin{equation}
\left[D_\nu,J^\nu\right]=0\; ,
\end{equation}
and the gauge condition is:
\begin{equation}
\partial_\mu A^\mu=0\; .
\label{eq:GC}
\end{equation}
Making use of the gauge condition, the Yang-Mills equation can be
rewritten as:
\begin{equation}
\square A^\nu=J^\nu+ig\left[A_\mu,F^{\mu\nu}+\partial^\mu A^\nu\right]\; .
\label{eq:YM-1}
\end{equation}

Let us first start with the terms $A^\mu_1$ and $A^\mu_2$, which are
linear in the hard sources. Since the commutator
$\left[A_\mu,F^{\mu\nu}+\partial^\mu A^\nu\right]$ is at least
quadratic in the sources, we can drop it at this order, and we have:
\begin{equation}
\square A^\mu_{1,2}=J^\mu_{1,2}\; ,
\label{eq:order1}
\end{equation}
where $J^\mu_{1,2}$ are the color currents associated to the
individual nuclei. If we assume that the nucleus $1$ is moving in the
$+z$ direction at the speed of light, while the nucleus $2$ is moving
in the $-z$ direction, the currents are simply:
\begin{eqnarray}
&&J^\mu_{1,a}=g\delta^{\mu +}\delta(x^-)\rho_{1,a}(\x_\perp)\; ,\nonumber\\
&&J^\mu_{2,a}=g\delta^{\mu -}\delta(x^+)\rho_{2,a}(\x_\perp)\; .
\end{eqnarray}
 The index $a$ is the color index carried by the classical sources and
currents.  This means in particular that at this order, the sources do
not have any transverse components, i.e. that they describe recoilless
objects.

Eq.~(\ref{eq:order1}) then simply becomes a Poisson equation in the
transverse plane, and the only non-zero components of $A^\mu_1$ and
$A^\mu_2$ can be written formally as:
\begin{eqnarray}
&&
A^+_{1,a}(x)
=-g\delta(x^-)\frac{1}{{\boldsymbol\nabla}_\perp^2}\rho_{1,a}(\x_\perp)\; ,\nonumber\\
&&A^-_{2,a}(x)
=-g\delta(x^+)\frac{1}{{\boldsymbol\nabla}_\perp^2}\rho_{2,a}(\x_\perp)\; .
\end{eqnarray}
In fact, this solution will be needed in momentum space when we
include it in the pair production amplitude:
\begin{eqnarray}
&&
A^+_{1,a}(k)=2\pi g\delta(k^-)\frac{1}{\k_\perp^2}\rho_{1,a}(\k_\perp)\; ,\nonumber\\
&&A^-_{2,a}(k)=2\pi g\delta(k^+)\frac{1}{\k_\perp^2}\rho_{2,a}(\k_\perp)\; .
\end{eqnarray}
In these formulas, $k$ can be seen as the momentum flowing from the
hard source to the quark line on which we insert the classical field. 

Let us now determine the contribution $A^\mu_{12}$ to the classical
field. If we isolate the terms of Eq.~(\ref{eq:YM-1}) that are
quadratic in the hard sources, we have:
\begin{equation}
\square A^\nu_{12}=J^\nu_{12}
+ig\left[
A_{\mu 1}+A_{\mu 2},F_1^{\mu\nu}+F_2^{\mu\nu}+\partial^\mu(A_1^\nu+A_2^\nu)
\right]\; ,
\end{equation}
where $J^\mu_{12}$ is the correction of order ${\cal O}(\rho_1\rho_2)$
to the color current. It can be determined by the current conservation
which, including terms that are quadratic in the sources, reads:
\begin{equation}
\partial_\nu J_1^\nu+\partial_\nu J^\nu_2+\partial_\nu J^\nu_{12}
-ig\left[A_{\nu 1}+A_{\nu 2},J^\nu_1+J^\nu_2\right]=0\; .
\end{equation}
The current conservation at linear order in the sources tells us that
$\partial_\nu J^\nu_{1,2}=0$, so that we have simply:
\begin{equation}
\partial_\nu J^\nu_{12}=ig\left[A_{\nu 1}+A_{\nu 2},
\square(A_{\nu 1}+A_{\nu 2})
\right]\; .
\label{eq:J12}
\end{equation}
 In addition to this equation, we know the following properties about
$J^\mu_{12}$:
\begin{eqnarray}
&&J^+_{12}\propto \delta(x^-)\theta(x^+)\; ,\nonumber\\
&&J^-_{12}\propto \delta(x^+)\theta(x^-)\; ,\nonumber\\
&&J^i_{12}=0\; .
\label{eq:J12-constrains}
\end{eqnarray}
The first two conditions mean that the $+$ (resp. $-$) component of
the current has to move along with the nucleus going in the $+$
(resp. $-$) direction, and that effects of the second nucleus cannot
start before the nuclei actually collide (hence the step
functions). The third condition is simply a statement that the
classical sources do not recoil.  Eq.~(\ref{eq:J12}) can be made more
explicit by writing\footnote{We denote
$\partial^+\equiv\partial/\partial x^-$ and
$\partial^-\equiv\partial/\partial x^+$.}:
\begin{equation}
\partial^+ J^-_{12}+\partial^- J^+_{12}=ig\left[T_a,T_b\right]\left\{
A^+_{1,a} \square A^-_{2,b}
-A^-_{2,b}\square A^+_{1,a}
\right\}\; ,
\label{eq:J12-1}
\end{equation}
where $T_{a,b}$ are color matrices. Moreover, the current $J^+_{12}$,
which follows the motion of nucleus $1$, must depend locally on the
source $\rho_1$ and receive a non-local correction from the second
nucleus. This means that in eq.~(\ref{eq:J12-1}) the term in
$A^+_{1,a} \square A^-_{2,b}$ (which is non-local in $\rho_1$ and
local in $\rho_2$) must go into $J^-_{12}$, and vice versa. At this
point, we can write formally:
\begin{eqnarray}
&&J^+_{12}=-ig\left[T_a,T_b\right]\frac{1}{\partial^-} A^-_{2,b}\square A^+_{1,a}\; ,\nonumber\\
&&J^-_{12}=ig\left[T_a,T_b\right]\frac{1}{\partial^+} A^+_{1,a} \square A^-_{2,b}\; .
\end{eqnarray}
Note that the inverses $1/\partial^\pm$, which are a priori not
uniquely defined, are made unambiguous by the step functions in
eqs.~(\ref{eq:J12-constrains}).

If we now realize that the only non-zero components of the first order
strength tensor are:
\begin{eqnarray}
&&F^{i+}_1=-F^{+i}_1=\partial^i A^+_1\; ,\nonumber\\
&&F^{i-}_2=-F^{-i}_2=\partial^i A^-_2\; ,
\end{eqnarray}
the evolution equations for $A^\mu_{12}$ read:
\begin{eqnarray}
&&\square A^+_{12}=J^+_{12}
+ig\left[A^-_2,\partial^+A^+_1\right]\; ,\nonumber\\
&&\square A^-_{12}=J^-_{12}
+ig\left[A^+_1,\partial^-A^-_2\right]\; ,\nonumber\\
&&\square A^i_{12}=-ig\left[A^+_1,\partial^iA^-_2\right]
-ig\left[A^-_2,\partial^iA^+_1\right]\; .
\end{eqnarray}
Using the explicit form of the current $J_{12}$, we can solve this as
follows:
\begin{eqnarray}
&&A^+_{12}=-ig\left[T_a,T_b\right]
\frac{1}{\square}\left\{
\left(\square A^+_{1,a}\right)\frac{1}{\partial^-}A^-_{2,b}
+\left(\partial^+ A^+_{1,a}\right)A^-_{2,b}
\right\}\; ,\nonumber\\
&&A^-_{12}=ig\left[T_a,T_b\right]
\frac{1}{\square}\left\{
\left(\square A^-_{2,b}\right)\frac{1}{\partial^+}A^+_{1,a}
+\left(\partial^- A^-_{2,b}\right)A^+_{1,a}
\right\}\; ,\nonumber\\
&&A^i_{12}=ig\left[T_a,T_b\right]\frac{1}{\square}\left\{
\left(\partial^i A^+_{1,a}\right)A^-_{2,b}
-\left(\partial^i A^-_{2,b}\right)A^+_{1,a}
\right\}\; .
\end{eqnarray}
Going to momentum space is trivial and leads to:
\begin{eqnarray}
&&A^+_{12}(k)=-ig\frac{\left[T_a,T_b\right]}{-k^2}\int\frac{d^4k_1}{(2\pi)^4}
\left\{
ik_1^+-\frac{k_1^2}{ik_2^-}
\right\}A^+_{1,a}(k_1)A^-_{2,b}(k_2)\; ,\nonumber\\
&&A^-_{12}(k)=ig\frac{\left[T_a,T_b\right]}{-k^2}\int\frac{d^4k_1}{(2\pi)^4}
\left\{
ik_2^--\frac{k_2^2}{ik_1^+}
\right\}A^+_{1,a}(k_1)A^-_{2,b}(k_2)\; ,\nonumber\\
&&A^i_{12}(k)=ig\frac{\left[T_a,T_b\right]}{-k^2}\int\frac{d^4k_1}{(2\pi)^4}
\left\{
ik_2^i-ik_1^i
\right\}A^+_{1,a}(k_1)A^-_{2,b}(k_2)\; ,
\end{eqnarray}
where we use the shorthand $k_2\equiv k-k_1$. One can note that the
order ${\cal O}(\rho_1\rho_2)$ correction to the classical gauge field
is proportional to a commutator of color matrices. This was to be
expected as it is known that in an Abelian gauge theory there would
not be such correction: the gauge field would simply be the sum of the
gauge fields created by the individual nuclei.

\section{Calculation of the traces}
\label{sec:traces}
In this appendix, we provide the result of the calculation of ${\rm
Tr}\,\left|m_{ab}^{-+}\right|^2$. In order to make this computation
more compact, let us first introduce the following
4-vectors:
\begin{eqnarray}
&&a\equiv (a^+\!=\!0,a^-\!=\!0,\a_\perp\!=\!\q_\perp-\k_{1\perp})\; ,\nonumber\\
&&b\equiv (b^+\!=\!0,b^-\!=\!0,\b_\perp\!=\!\k_{1\perp}-\p_\perp)\; ,\nonumber\\
&&c\equiv\left(c^+\!=\!p^+\!+\!q^+\!-\!\frac{\k_{1\perp}^2}{p^-\!+\!q^-},
c^-\!=\!\frac{\k_{2\perp}^2}{p^+\!+\!q^+}\!-\!p^-\!-\!q^-,
\cc_\perp\!=\!\k_{2\perp}\!-\!\k_{1\perp}
\right)\; ,\nonumber\\
&&
\end{eqnarray}
and the following notations for the denominators:
\begin{eqnarray}
&&2q^-p^++(\q_\perp-\k_{1\perp})^2+m^2\equiv m^2-\hat{t}\; ,\nonumber\\
&&2q^+p^-+(\p_\perp-\k_{1\perp})^2+m^2\equiv m^2-\hat{u}\; ,\nonumber\\
&&(p+q)^2\equiv \hat{s}\; ,
\end{eqnarray}
where $\hat{s}$, $\hat{t}$ and $\hat{u}$ are the standard Mandelstam
variables for the $gg\to q\bar{q}$ subprocess. We can rewrite the
amplitude as follows:
\begin{eqnarray}
&&m_{ab}^{-+}(k_1,k_2;\q,\p)=\nonumber\\
&&\quad=ig^2 \overline{u}(\q)\Bigg\{
t_a t_b
\left[
\frac{\gamma^-(m+\sla)\gamma^+}{m^2-\hat{t}}
+\frac{\slc}{\hat{s}}
\right]
+
t_b t_a
\left[
\frac{\gamma^+(m+\slb)\gamma^-}{m^2-\hat{u}}
-\frac{\slc}{\hat{s}}
\right]
\Bigg\}v(\p)\; .\nonumber\\
&&
\end{eqnarray}
Depending on how terms are paired when squaring the amplitude, there
are two kinds of color traces:
\begin{eqnarray}
&&{\rm tr}_c\,(t_at_at_bt_b)=N_c C_{_{F}}^2\; ,\nonumber\\
&&{\rm tr}_c\,(t_at_bt_at_b)=-\frac{1}{2} C_{_{F}}\; ,
\end{eqnarray}
where $C_{_{F}}\equiv (N_c^2-1)/(2N_c)$ is the Casimir in the
fundamental representation of $SU(N_c)$. In the large $N_c$ limit, the
first trace scales like $N_c^3$, while the second trace scales only
like $N_c$. We can therefore write the trace of the squared amplitude
as:
\begin{equation}
{\rm Tr}\,\left(\left|
m_{ab}^{-+}(k_1,k_2;\q,\p)
\right|^2\right)=g^4 C_{_{F}}\left[
N_c C_{_{F}} T_3-\frac{1}{2} T_1
\right]\; ,
\end{equation}
where we denote:
\begin{eqnarray}
&&T_3\equiv{\rm tr}\,\Bigg\{
(\slq+m)\left[
\frac{\gamma^-(m+\sla)\gamma^+}{m^2-\hat{t}}
+\frac{\slc}{\hat{s}}
\right]
(\slp-m)\left[
\frac{\gamma^+(m+\sla)\gamma^-}{m^2-\hat{t}}
+\frac{\slc}{\hat{s}}
\right]\nonumber\\
&&\qquad
+
(\slq+m)\left[
\frac{\gamma^+(m+\slb)\gamma^-}{m^2-\hat{u}}
-\frac{\slc}{\hat{s}}
\right]
(\slp-m)\left[
\frac{\gamma^-(m+\slb)\gamma^+}{m^2-\hat{u}}
-\frac{\slc}{\hat{s}}
\right]
\Bigg\}
\end{eqnarray}
and
\begin{eqnarray}
&&T_1\equiv{\rm tr}\,\Bigg\{
(\slq+m)\left[
\frac{\gamma^-(m+\sla)\gamma^+}{m^2-\hat{t}}
+\frac{\slc}{\hat{s}}
\right](\slp-m)\left[
\frac{\gamma^-(m+\slb)\gamma^+}{m^2-\hat{u}}
-\frac{\slc}{\hat{s}}
\right]\nonumber\\
&&\qquad+
(\slq+m)\left[
\frac{\gamma^+(m+\slb)\gamma^-}{m^2-\hat{u}}
-\frac{\slc}{\hat{s}}
\right](\slp-m)\left[
\frac{\gamma^+(m+\sla)\gamma^-}{m^2-\hat{t}}
+\frac{\slc}{\hat{s}}
\right]
\Bigg\}\; ,
\end{eqnarray}
where the subscripts $1,3$ refer to the order of the corresponding
terms when $N_c\to \infty$. It is in fact a bit simpler to write
\begin{equation}
{\rm Tr}\,\left(\left|
m_{ab}^{-+}(k_1,k_2;\q,\p)
\right|^2\right)=\frac{1}{2}g^4 C_{_{F}}\left[ 
N_c^2 T_3-T_1^\prime
\right]\; ,
\end{equation}
with $T_1^\prime\equiv T_1+T_3$, because it turns out that
$T_1^\prime$ has a more compact expression that $T_1$. A direct
calculation using {\sc form}\cite{Verma1} leads after some
rearrangement of the terms to the following expressions:
\begin{eqnarray}
&&T_1^\prime =16\Bigg\{
        \frac{(m^2-a^2)p^+q^-}{(m^2-\hat{t})^2}
        +
        \frac{(m^2-b^2)p^-q^+}{(m^2-\hat{u})^2}
        \nonumber\\
&&
        +
        \frac{(a\!\cdot\! b\!-\!m^2)(\q_\perp\!\cdot\!\p_\perp\!-\!m^2)
        \!-\!m^2(\q_\perp\!-\!\p_\perp)^2
        \!+\!(p\!\cdot\! a) (q\!\cdot\! b)\!+\!(p\!\cdot\! b)( q\!\cdot\! a)\!-\!2m^4}
{(m^2-\hat{t})(m^2-\hat{u})}\Bigg\}\nonumber\\
&&
\end{eqnarray}
and
\begin{eqnarray}
&&T_3=\frac{16(m^2-a^2)p^+q^-}{(m^2-\hat{t})^2}
        +
        \frac{16(m^2-b^2)p^-q^+}{(m^2-\hat{u})^2}
        -
        \frac{16(q\!\cdot\! c)^2+4c^2\hat{s}}{\hat{s}^2}
\nonumber\\
&&
        \qquad+
        \frac{8}{\hat{s}(m^2-\hat{t})}
        \Big[
        (a\!\cdot\! c)(p^+q^--p^-q^++\hat{s}/2)
\nonumber\\
&&\qquad\qquad
        +(2p^+c^--\p_\perp\!\cdot\! \cc_\perp)(m^2-q\!\cdot\! a)
        -(2q^-c^+-\q_\perp\!\cdot\! \cc_\perp)(m^2+p\!\cdot\! a)
        \Big]
\nonumber\\
&&
        \qquad+
        \frac{8}{\hat{s}(m^2-\hat{u})}
        \Big[
        (b\!\cdot\! c)(p^+q^--p^-q^+-\hat{s}/2)
\nonumber\\
&&\qquad\qquad
        +(2q^+c^--\q_\perp\!\cdot\!\cc_\perp)(m^2+p\!\cdot\! b)
        -(2p^-c^+-\p_\perp\!\cdot\!\cc_\perp)(m^2-q\!\cdot\! b)
        \Big]\; .\nonumber\\
&&
\end{eqnarray}
It is straightforward to verify that $T_1^\prime$ and $T_3$ vanish in
the limits $\k_{1\perp}\to 0$ and $\k_{1\perp}\to \p_\perp+\q_\perp$.

One can also verify that our expressions, derived by solving
perturbatively the classical Yang-Mills equation, are strictly
equivalent to the formulas obtained in the framework of
$\k_\perp$-factorization by Collins and Ellis \cite{ColliE1} (also
derived independently by Catani et al in \cite{CatanCH1}). In order to
perform this comparison, one needs the following dictionnary (our
notations are on the left, and theirs are on the right):
\begin{eqnarray}
q  &\Longleftrightarrow & p_3\; ,\nonumber\\
p  &\Longleftrightarrow & p_4\; ,
\end{eqnarray}
(the notation for the momenta $k_{1,2}$ of the incoming gluons is the
same). The relationship between our amplitude squared and the formulas
given in eqs.~(5.7) and (5.9) of \cite{ColliE1} is then summarized by:
\begin{eqnarray}
&&T_1^\prime = 8 Y_2\; ,\nonumber\\
&&T_3=4 (Y_1+Y_2)\; .
\end{eqnarray}
However, we do not agree with the amplitude squared given in the
appendix of \cite{LevinRSS1}. A private communication from Levin
acknowledged the presence of a mistake in the formulas quoted in
\cite{LevinRSS1}, but the corrected expressions were unavailable to us
in order to compare them with our result.

\bibliographystyle{unsrt}

\end{document}